\documentclass[]{spie}  

\usepackage{amsmath,amsfonts,amssymb}
\usepackage{graphicx}
\usepackage[colorlinks=true, allcolors=blue]{hyperref}

\title{The Visible Integral-field Spectrograph eXtreme (VIS-X): high-resolution spectroscopy with MagAO-X }

\author[a]{Sebastiaan Y. Haffert}
\author[a]{Jared R. Males}
\author[a]{Laird M. Close}
\author[a,b,c]{Kyle Van Gorkom}
\author[a]{Joseph D. Long}
\author[a,b]{Alexander D. Hedglen}
\author[a,b,d,e]{Olivier Guyon}
\author[a,b]{Lauren Schatz}
\author[a,b]{Maggie Kautz}
\author[a,b]{Jennifer Lumbres}
\author[a,b]{Alexander Rodack}
\author[a,b]{Justin M. Knight}

\affil[a]{University of Arizona, Steward Observatory, Tucson, Arizona, United States}
\affil[b]{Wyant College of Optical Science, University of Arizona, 1630 E University Blvd, Tucson, AZ 85719, USA}
\affil[c]{NASA Goddard Space Flight Center, Greenbelt, MD 20771, USA}
\affil[d]{Astrobiology Center, National Institutes of Natural Sciences, 2-21-1 Osawa, Mitaka, Tokyo, JAPAN}
\affil[e]{National Astronomical Observatory of Japan, Subaru Telescope, National Institutes of Natural Sciences, Hilo, HI 96720, USA}

\authorinfo{Further author information: (Send correspondence to S.Y.H.)\\S.Y.H.: E-mail: shaffert@arizona.edu\\ S.Y.H. is a NASA Hubble fellow}

\begin{document} 
\maketitle

\begin{abstract}
MagAO-X system is a new adaptive optics for the Magellan Clay 6.5m telescope. MagAO-X has been designed to provide extreme adaptive optics (ExAO) performance in the visible. VIS-X is an integral-field spectrograph specifically designed for MagAO-X, and it will cover the optical spectral range (450 – 900 nm) at high-spectral (R=15.000) and high-spatial resolution (7 mas spaxels) over a 0.525 arsecond field of view. VIS-X will be used to observe accreting protoplanets such as PDS70 b \& c. End-to-end simulations show that the combination of MagAO-X with VIS-X is 100 times more sensitive to accreting protoplanets than any other instrument to date. VIS-X can resolve the planetary accretion lines, and therefore constrain the accretion process. The instrument is scheduled to have its first light in Fall 2021. We will show the lab measurements to characterize the spectrograph and its post-processing performance.
\end{abstract}


\section{INTRODUCTION}
Young stars form in dense clouds that collapse under their own gravity. The angular momentum that is present at the beginning of this process forces the material to clump together into a disk around the young star. Planets form inside these disks either through instabilities in the disk that cause the formation of dens clumps or through pebble accretion. Multiple pathways to the formation of planets have been proposed and there is no clear answer yet on what each process's contribution is. A thorough understanding of the formation and evolution of exoplanets will allow us to gain an understanding of not only the origin of Earth, but possibly life. To this end we will need to characterize exoplanets in all stages of their evolution, from young to old. However, the most common characterization technique utilizes the transit method which suffers from astrophysical limitations such as constraints on the orbital geometry or astrophysical noise due to e.g. star spots or circumstellar material. Direct imaging plays an important role to overcome these observational limitations. For both the young and old systems, the influence of the star, and possible circum-stellar material, can be significantly reduced by spatially resolving the planet from its environment, allowing detailed characterization of the planet.

The past few years has seen a large step in the capabilities of direct imaging instruments. Instruments such as SPHERE \cite{beuzit2019sphere} or MagAO \cite{close2014into} have observed the environment around many young stars in search of giant proto-planets; planets that are still accreting material from their birth environment. Young proto-planets sweep up the gas and dust within the proto-planetary disk. The accretion of gas releases a large amount of energy when the gas falls onto the planet, or its circum-planetary disks. Most of this energy is released by specific line emission such as the hydrogen emission lines \cite{aoyama2020spectral}. These signatures are therefore one of the strongest signposts of accreting gas giants. Several instruments contain sets of narrowband imaging filters that image the emission lines and the nearby continuum \cite{close2014discovery}. Recent observations have shown that medium to high-resolution spectroscopy is ideal to observe accreting proto-planets \cite{haffert2019pds70, xie2020searching}. However, no direct imaging instrument has the capability of visible light high-resolution integral-field spectroscopy.

The Magellan Adaptive Optics eXtreme (MagAO-X) system is a new adaptive optics for the Magellan Clay 6.5m telescope at Las Campanas Observatory (LCO). MagAO-X has been designed to provide extreme adaptive optics (ExAO) performance in the visible. It will ultimately deliver Strehl ratios of 90\% at 0.9 $\mu$m and nearly 80\% at H$\alpha$ (Males et al., 2018). The performance of MagAO-X in the visible is comparable to what other direct imaging instruments achieve in H or K-band, making MagAO-X the ideal instrument to push exoplanet characterization to the visible range. However, MagAO-X does not have any spectroscopic capability. The Visible Integral-field Spectrograph eXtreme (VIS-X) is a spectrograph for MagAO-X that will cover the optical spectral range at high-spectral and high-spatial resolution. Sections 2 will expand on the primary science case and the instrumental requirements. Then Section 3 will discuss the instrument design and the first lab results.

\section{Primary science case for VIS-X: Accreting proto-planets}
The accretion process of sub-stellar companions is a key part of the information that can be used to discriminate between the different formation processes. The ability to accrete gas at all and the actual mass accretion rate will allow us to discriminate between formation pathways \cite{stamatellos2015properties}. Gas accretion on massive planets is thought to be a very energetic process, and the emitted accretion luminosity can become comparable to the total internal luminosity of the planet \cite{mordasini2017characterization}. Therefore, visible-light High-Contrast Imaging (HCI) is a promising approach to detect these young protoplanets, because it provides access to strong accretion tracers such as H$\alpha$. The huge potential of H$\alpha$ imaging has been demonstrated by recent detections of actively accreting companions with HST/WFC (Zhou et al. 2014), Magellan/MagAO \cite{close2014discovery, wu2017alma, wagner2018pds70}, and more recently VLT/MUSE \cite{haffert2019pds70, xie2020searching, eriksson2020strong }.

Haffert et al. 2019 show that integral-field spectroscopy is a very powerful technique to unambiguously detect proto-planets. Conventional high-contrast imaging techniques, such as Angular Differential Imaging (ADI) \cite{marois2006adi}, often lead to point-like features that are caused by either residual instrumental artifacts or due to the presence of non-symmetric circumstellar disks \cite{ligi2018investigation}. With high-resolution integral-field spectroscopy we can eliminate the star and the circum-stellar disk by removing a scaled stellar spectrum from each spatial pixel (spaxel). The signal from the circum-stellar disk is also removed during this procedure because the light from the disk in the visible range consists mainly of reflected star light and is therefore identical to the stellar spectrum. 

The emission lines from accreting planets are very narrow \cite{aoyama2020spectral}. Marleau et al. in prep show that the H$\alpha$ emission line starts to be resolved around a resolving power of 15,000 (20 km/s). This implies that optimal SNR can be achieved when the resolving power of the instrument is 15,000. If the resolving power is increased even further the light from the proto-planet will be smeared out over more detector pixels, which will increase the detector noise contribution. The signal to noise (SNR) of the H$\alpha$ measurement is,

\begin{equation}
    \mathrm{SNR} = \frac{T F_{\mathrm{H\alpha},P}}{\sqrt{T (F_{\mathrm{Phot},S} + F_{\mathrm{H\alpha},S}) C(\theta_P) + N\sigma_D^2 }}.
\end{equation}
Here $T$ is the throughput from the top of the atmosphere to the camera, $F_{\mathrm{H\alpha},P}$ and $F_{\mathrm{H\alpha},S}$ are the H$\alpha$ flux of the planet and star, $F_{\mathrm{Phot},S}$ is the stellar photospheric emission, $C(\theta_P)$ the contrast of the observations at angular separation $\theta_P$, $N$ is the number of pixels that are used to sample an unresolved emission line and $\sigma_D$ is the detector noise (read noise + dark current). Under favourable conditions the H$\alpha$ line of the star is separated from the H$\alpha$ line of the planet due to an intrinsic radial velocity difference. At high enough spectral resolution the stellar H$\alpha$ flux will not contribute at all at the velocity position of the planet. The flux of the stellar continuum is proportional to the bandwidth of a single spectral slice. Taking this into account we arrive at the following SNR for the H$\alpha$ detection,
\begin{equation}
    \mathrm{SNR} = \frac{T F_{\mathrm{H\alpha},P}}{\sqrt{T 
    \langle F_{\mathrm{Phot}, S} \rangle \delta \lambda C(\theta_P) + N\sigma_D^2 }}.
\end{equation}
Here the photospheric flux of the star has been replaced by the average flux density times the bandwidth of a single spectral channel. The bandwidth $\delta \lambda = \lambda / R$, where $R$ is the resolving power of the spectrograph. In the regions where photon noise dominates, the SNR is
\begin{equation}
    \mathrm{SNR} = \frac{F_{\mathrm{H\alpha},P} \sqrt{TR}}{\sqrt{
    \langle F_{\mathrm{Phot}, S} \rangle \lambda C(\theta_P)}}.
\end{equation}
This shows that the expected SNR scales with $\sqrt{R}$, under the assumption that the spectral line is unresolved. A high-resolution spectrograph at $R=15,000$ can have a large gain in sensitivity compared to narrowband imaging ($R=100$ for MagAO, and $R=120$ or $R\sim650$ for the broadband and narrowband H$\alpha$ filters in SPHERE, respectively). A rough order of magnitude in SNR improvement is $\sqrt{15000 / 100}\approx12.3$. This shows that significant gains can be made by using high-resolution integral field spectroscopy for accreting proto-planets.

\subsection{Post-processing gain}
There are several aspects of high-resolution integral-field spectroscopy for emission line imaging. The first and foremost advantage of HRS, is that it is easier to disentangle light from the emission lines from the neighboring continuum. Only two images are taken with the classic approach of dual band imaging. In post-processing, the continuum image is magnified by $\lambda_1 / \lambda_2$ to take care of the chromatic scaling due to diffraction. For accurate subtraction of one channel from the other, the flux total flux has to be scaled. This is usually achieved by measuring the total flux within the Airy core. After the flux correction, the images can be subtracted from each other,
\begin{equation}
\delta I = I(\lambda_1) - a I(\lambda_2) - b.
\end{equation}
Here $I(\lambda_i)$ is the observed image at spectral channel $i$, $a$ is the linear scale parameter, and $b$ is the background. This approach can be used to gain in contrast, however the gain is limited because this model assumes that the diffraction pattern and speckles can be modeled by a global linear scaling. This approach would work for a system without any wavefront aberration. However, any amount of wavefront aberration will introduce non-linear chromatic behavior in the focal plane speckles. This is especially true when amplitude errors are also present. The pupil plane electric field can be represented by,
\begin{equation}
E_p = A(1+g) e^{i \frac{2\pi}{\lambda} \delta}.
\end{equation}
Here $E_p$ is the pupil plane intensity, $A$ the pupil function, $g$ the amplitude aberrations and $\delta$ the wavefront error. For high-contrast imaging instruments $\delta$ is usually small and the exponential can be expanded into its Taylor series,
\begin{equation}
E_p = A(1+g) e^{i \frac{2\pi}{\lambda} \delta} \approx A(1+g)\left(1 + \frac{2\pi i}{\lambda} \right).
\end{equation}
The propagation from the input pupil plane through the optical system to the final science focal plane is represented by the linear operator $C$. With this operator in hand, the focal plane electric field is $E_f = CE_p$. This representation holds for any type of IFS implementation, such as micro-lens array based IFS's or dual band imagers. Detectors can not measure the electric field directly and measure the intensity. This means that the final intensity that we measure is,
\begin{equation}
I_p = |CP +CP\frac{2\pi i}{\lambda}|^2 = |CP|^2 +  |CP\frac{2\pi i}{\lambda}|^2 + 2\Re\left\{CP \left(CP \frac{2\pi i}{\lambda}\delta\right)^{\dagger}\right\}.
\end{equation}
From this it is clear that even in the small-aberration regime, there are terms that have a different chromatic behavior. The aberration-free term has no chromatic scaling except for the chromatic magnification due to diffraction. And the other two terms both scale differently with wavelength. In this example, the wavefront error was achromatic and the higher-order terms have been neglected. Such a simple example already shows why there is no global linear relation between two different spectral channels, and also explains why DBI will not remove all aberrations.

\begin{figure} [ht]
	\begin{center}
		\includegraphics[]{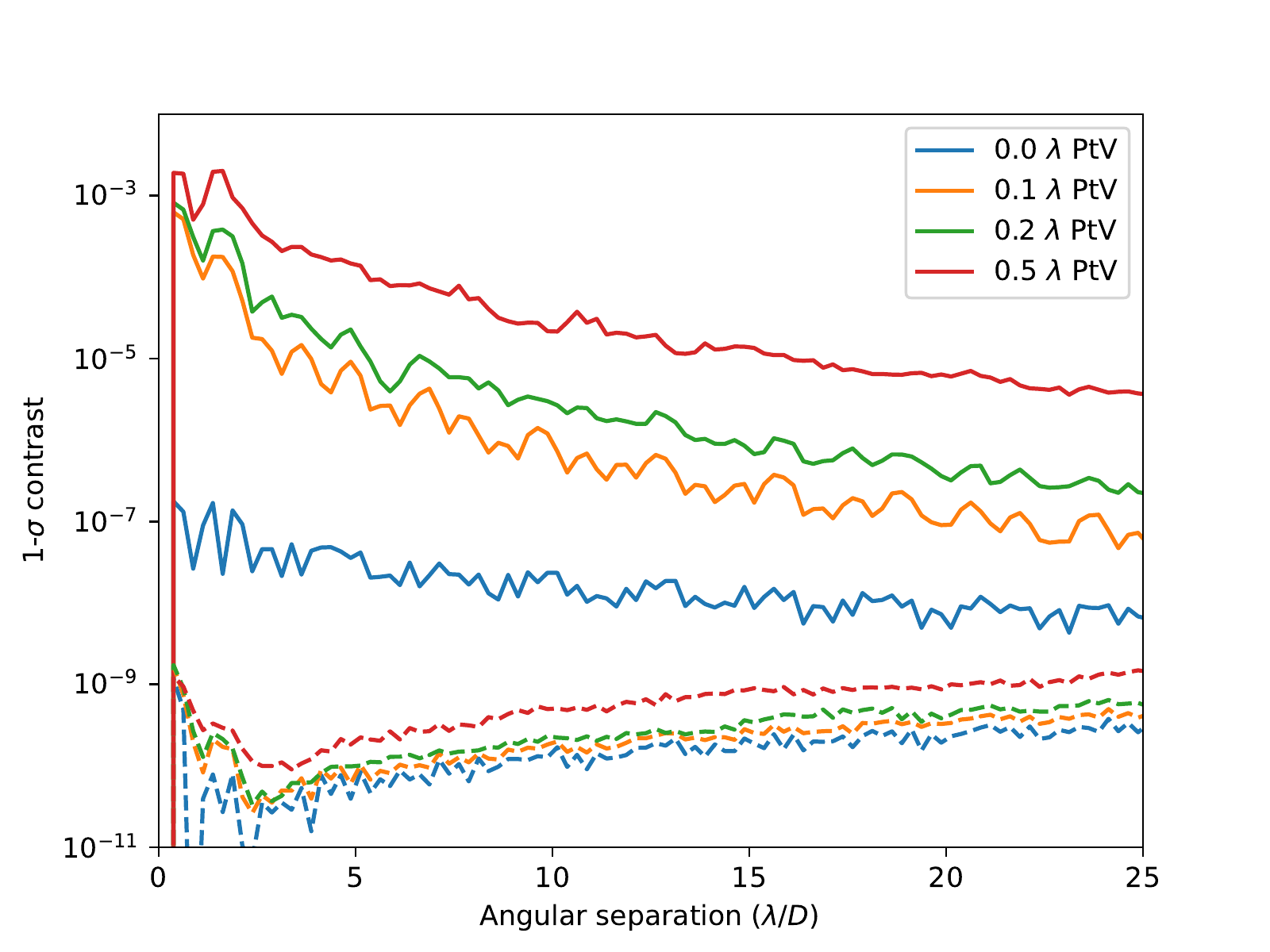}
	\end{center}
	\caption[example]
	{ \label{fig:example} 
		The radial 1-$\sigma$ contrast curve after post-processing. For the DBI method we applied the optimal scaling of the two spectral channels. Each color represents a different amount of wavefront aberration. The solid curves are the contrast curves after DBI post-processing and the dashed curves are the contrast curves for the IFS post-processing. For the DBI  method the contrast decreases when the wavefront error is increased, while for the IFS observations the residuals increases slightly and always stays well below the DBI residuals. When no wavefront errors are present the 1-$\sigma$ contrast for the DBI method is between $10^{-7}$ and $10^{-8}$.}
\end{figure} 

More spectral channels have to be measured to accurately model the chromatic behavior of the stellar speckles. A IFS provide such an opportunity. However, having many spectral channels is not the only requirement. Chromatic scaling due to diffraction happens on spectral resolving powers of $R=N$, with $R$ the resolving power and $N$ the field of view in units of $\lambda/D$. Typical instruments have a field of view of $\approx100 \lambda/D$. The instrument should have a $R\gg100$ to accurately measure the speckles. However, most if not all direct imaging IFS have low spectral resolving power, e.g. SPHERE-IFS has 50, CHARIS/SCEXAO has 20-80 and ALES at the LBT has 40. Speckle removal with higher-resolution spectroscopy has already proven itself to work better than DBI\cite{hoeijmakers2018atomic}.

The post-processing of HR-IFS data uses the assumption that diffraction and instrumental effects happen at low-spectral resolution and can therefore be modeled by low-order polynomials. This means that the spectrum at every pixel can described by a reference stellar spectrum multiplied by a low-order polynomial. There are many spectral channels available for each spatial pixel(spaxel) in the IFU, therefore a different polynomial model can be estimated for each spaxel. This is described by,
\begin{equation}
I_j(\lambda) = \sum_i a_{ij} \phi_i(\lambda) S(\lambda).
\end{equation}
A Linear Least-Squares (LLS) solution can be found for the polynomials coefficients because the model is linear in the coefficients. A good choice of low-order polynomials are Chebyshev or Legendre polynomials. These are orthogonal and create robust matrices for the matrix inversion step.

The results of the post-processing gain are shown in Figure \ref{fig:example}. The high-resolution observations are well below $10^{-8}$, which is more than sufficient to detect accreting proto-planets. The DBI mode can have strong residuals, depending on the exact chromatic behavior of the speckles. The DBI method is limited to $>10{-6}$ at the inner regions for smallest wavefront errors. This simulation only considered a single phase screen in the pupil. Real systems are expected to contain stronger and more complex chromatic behavior which will push the post-processed contrast to the $0.5\lambda$ curve.

\subsubsection{First end-to-end simulations}
\begin{figure} [ht]
	\begin{center}
		\includegraphics[width=\textwidth]{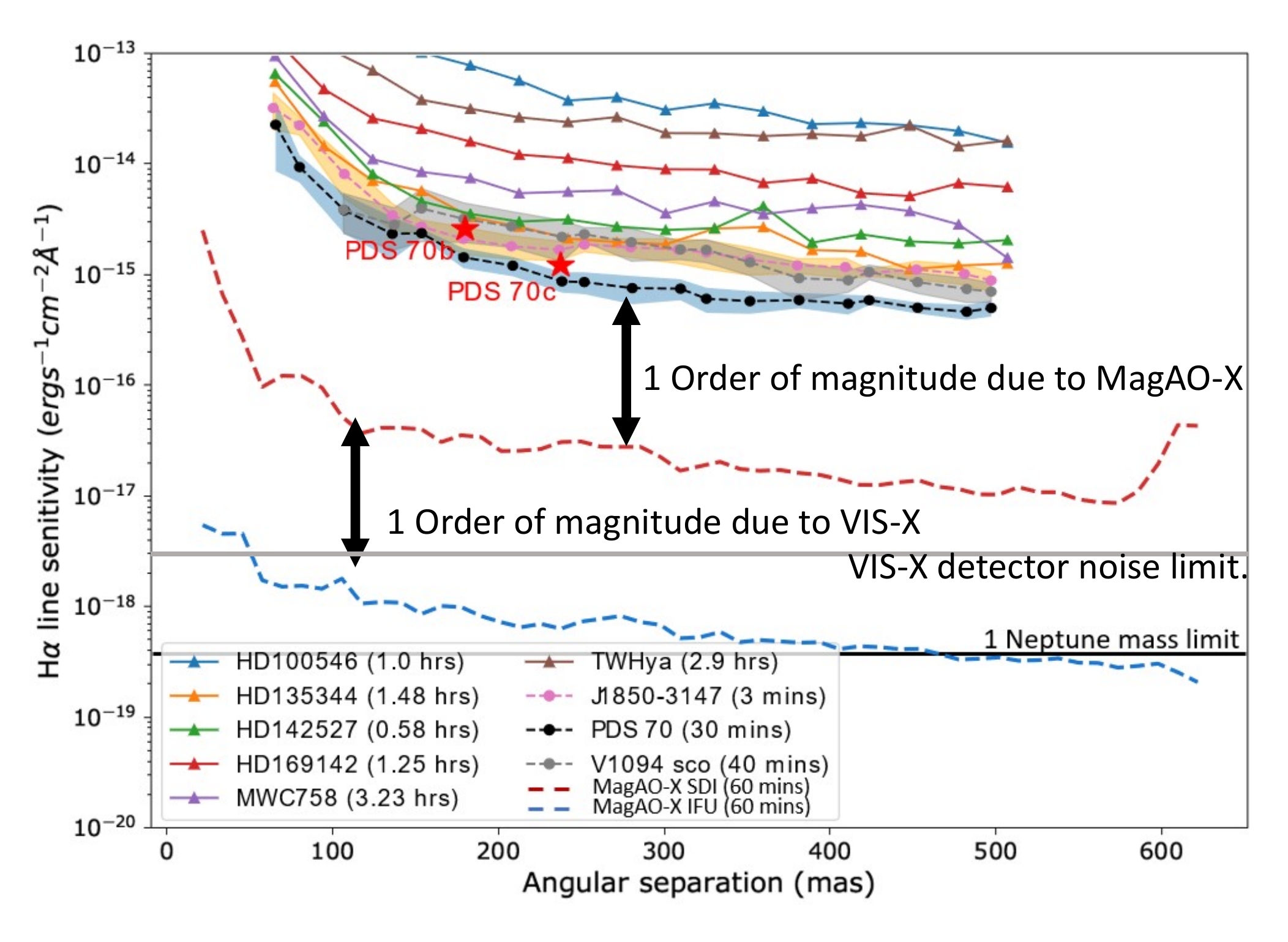}
	\end{center}
	\caption[example]
	{ \label{fig:sensitivity} 
		The radial 1-$\sigma$ contrast curve after post-processing. The red dased curve represents the sensitivity of MagAO-X with dual band imaging, while the blue dashed curve shows the sensitivity with VIS-X. Both curves are compared to actual observations from MUSE and SPHERE. This figure has been adapted from \cite{xie2020searching}}
\end{figure}

End-to-end simulations of an $R=15000$ H$\alpha$ spectrograph coupled to MagAO-X have been performed to investigate the gain in performance. We have simulated a system with 50 actuators across the pupil driven by a unmodulated pyramid wavefront sensor. The star itself was an 8th magnitude star. Median atmospheric conditions of the Las Campanas site have been assumed ($v=15m/s$, $r_0=0.16$ at $\lambda=550$ nm). A separate static phase screen with 50 nm rms has been used to simulate non-common path errors. The sensitivity curves are shown in Figure \ref{fig:sensitivity}. The derived contrast curves show that VIS-X will indeed add roughly a factor 10 to the sensitivity of MagAO-X. MagAO-X itself will already be more sensitive than any other instrument because the instrument has been optimized to operate as an xAO system in the visible. VIS-X shows the larges gain in the inner few $\lambda/D$, exactly where we expect to find the most planets \cite{close2020separation}. This demonstrates the benefit of high-resolution IFU observations of the H$\alpha$ emission line. Additionally, due to the higher resolution it may become possible to study the line shapes and derive the physical state of the accretion process.

\section{VIS-X DESIGN AND FIRST MEASUREMENTS}
Due to constraints on detector real estate there is a trade-off between spatial sampling, field-of-view, spectral bandwidth and spectral resolution. Maximizing the field of view and spectral resolving power requires a large format detector. In the past few years there has been a significant amount of progress of in the quality of back-illuminated CMOS detector technology. SONY released the imx455 sensor in 2019 with 9600x6422 pixels, a quantum efficiency close to 80 percent at H$\alpha$ and a peak efficiency close to 90 percent (at 550 nm) and a ~1 electron read noise at the highest gain setting. With a water-based cooling it is possible to reduce the dark current to below 0.001 electron/s/pixel. These properties make this an ideal sensor for visible integral-field spectroscopy.

\begin{figure} [ht]
	\begin{center}
		\includegraphics[width=\textwidth]{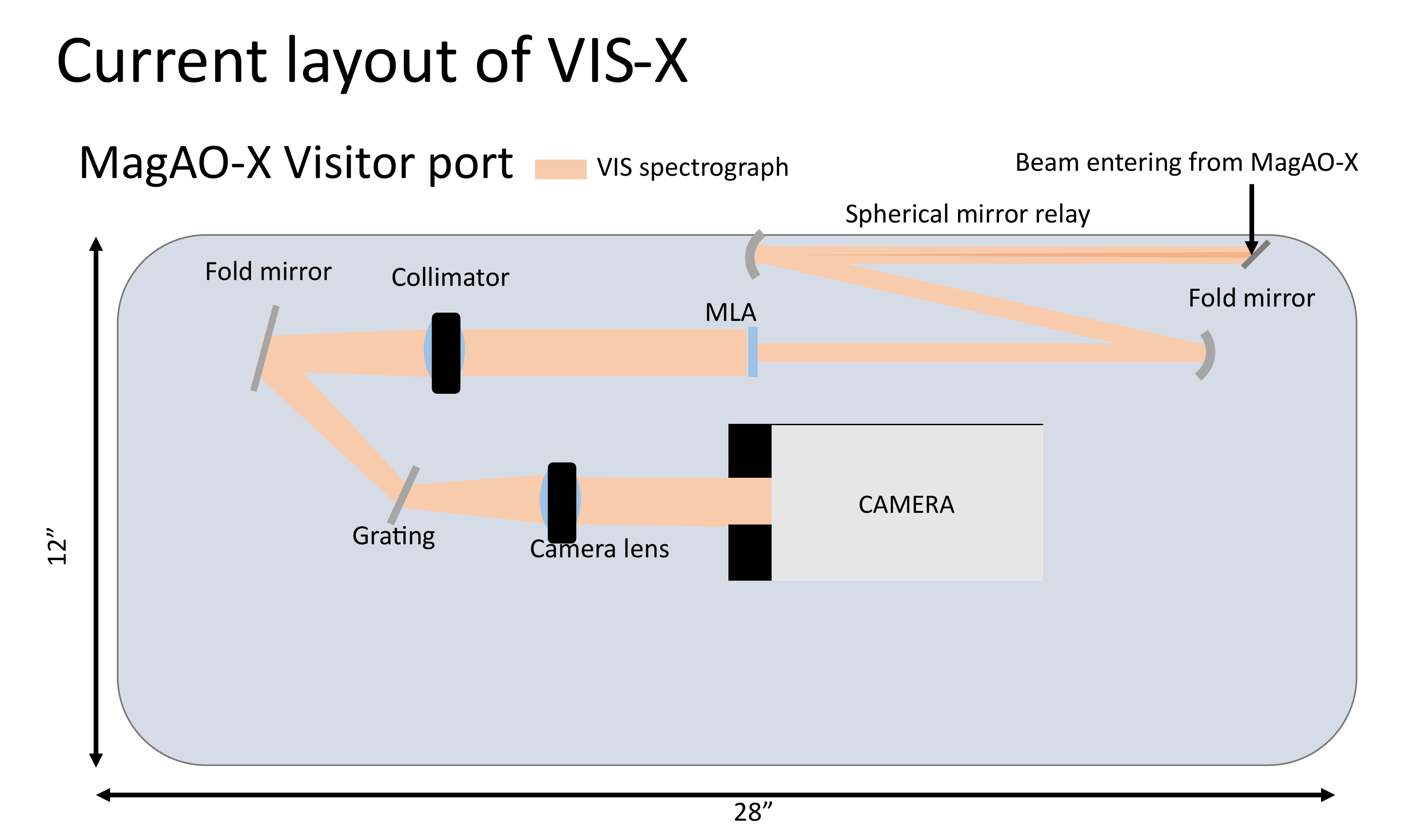}
	\end{center}
	\caption[example]
	{ \label{fig:layout} 
		A schematic drawing of VIS-X in the available space envelop of MagAO-X. The beam from MagAO-X enters from the top right and follows the path in the direction of the arrow. A spherical mirror-based relay will magnify the PSF onto an MLA. The MLA output is then dispersed by a first-order grating spectrograph.}
\end{figure} 

\begin{figure} [ht]
	\begin{center}
		\includegraphics[width=\textwidth]{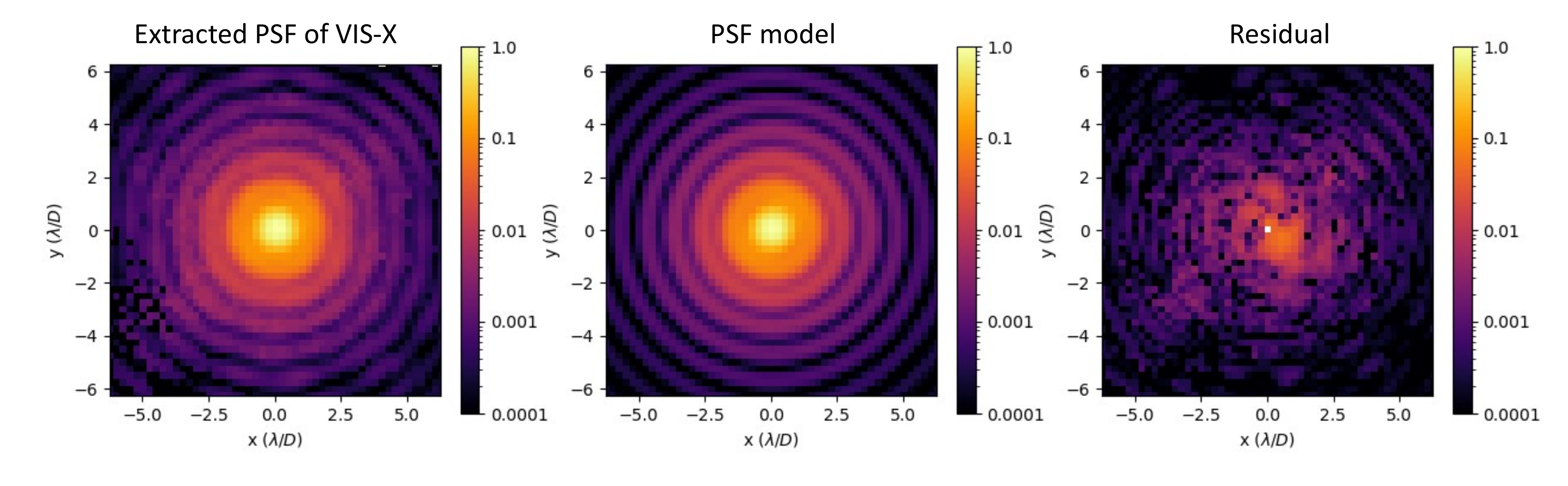}
	\end{center}
	\caption[example]
	{ \label{fig:extracted_psf} 
		The reconstructed PSF from VIS-X. There is a slight defocus visible (~$\lambda$/10 rms), and some extraction artifacts on the lower left due to the low SNR of the calibration files. The middle figure shows the fitted PSF model, and the right figure shows the residuals. The residuals are on the order of $\sim10^{-2}$.}
\end{figure}

With an internal UA seed grant, we developed a prototype micro-lens array (MLA) based integral-field spectrograph that can operate in a narrowband (6 nm) around H$\alpha$ using the new imx455 sensor. MLA based IFUs are in use in all direct imaging instrument and are considered a mature technology and therefore a low risk design. The prototype has been designed to deliver a resolving power of R$\sim15000$ at H$\alpha$, with a fixed spectral bandwidth of 5nm. The prototype has a limited field of view of 0.5” in diameter. Figure \ref{fig:layout} shows a schematic of the spectrograph within the available space envelop of MagAO-X. We use two spherical mirrors as an achromatic relay that magnifies the F/69 beam of MagAO-X to sample the PSF with 3 spaxels per $\lambda/D$ at H$\alpha$ with the MLA. This will keep us Nyquist sampled down to H$\gamma$ (434 nm). On-sky experience with the MUSE IFU at the VLT showed us that well sampled LSF’s are critical for accurate post-processing\cite{xie2020searching}. The relay itself has a theoretical wavefront rms of $\lambda/100$ because we are working with slow beams (F/69 to F/870). The performance is therefore mainly determined by the manufacturing quality of the relay mirrors. We found that $\lambda$/4 mirrors are fitting for our purpose, and lab measurements confirmed that there is no indication of degradation of the PSF after the relay, see Figure \ref{fig:extracted_psf} for an extracted PSF. The extracted PSF contains roughly $\lambda$/10 rms defocus which is well within the correction range of the NCPA DM of MagAO-X.

\begin{figure} [ht]
	\begin{center}
		\includegraphics[width=\textwidth]{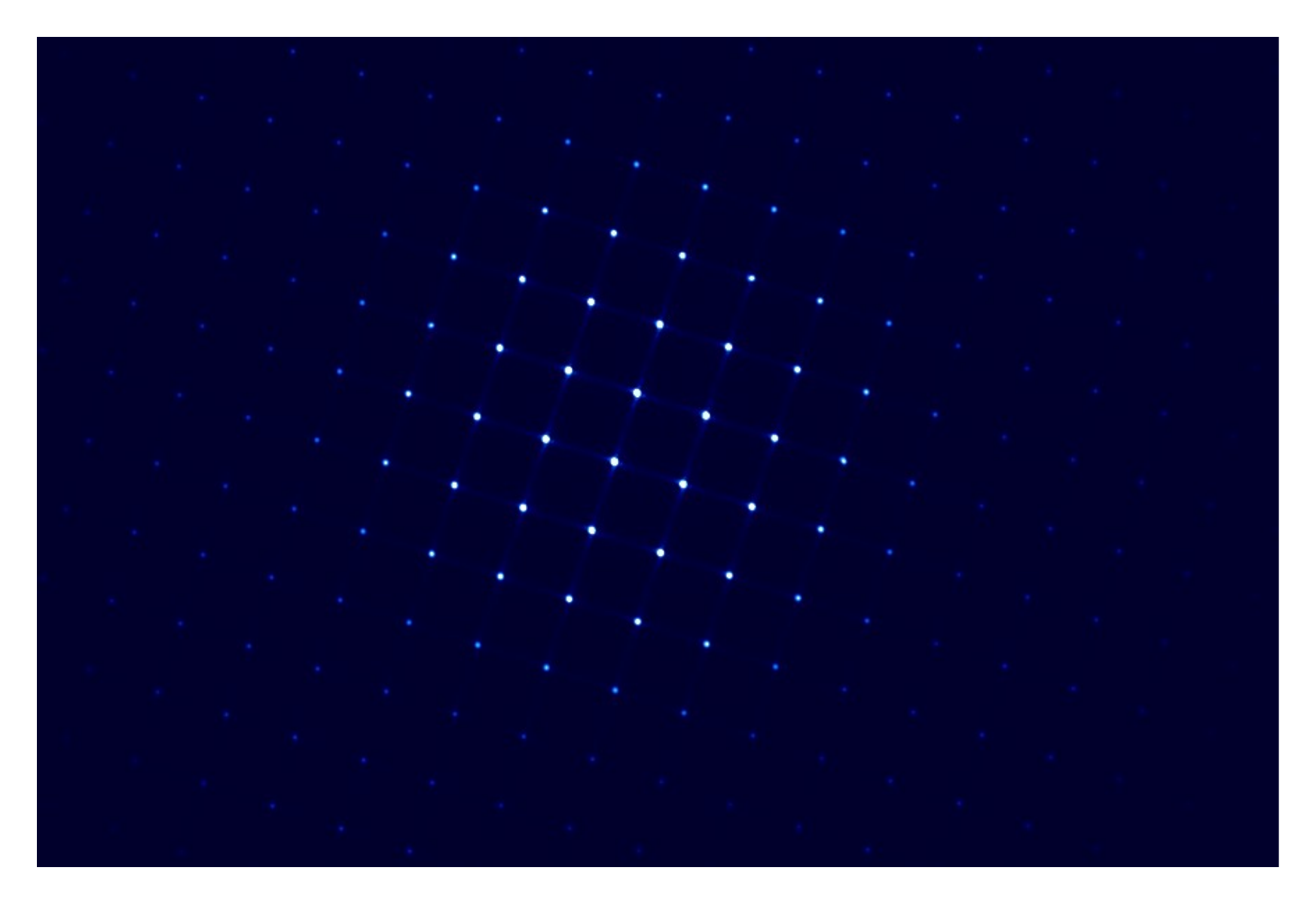}
	\end{center}
	\caption[example]
	{ \label{fig:psflets} 
		The PSFs of several spaxels. The area that is shown here is part of the center of the IFU. All spaxels are well separated from each other and show diffraction-limited image quality.}
\end{figure}

The PSF is sampled by a micro-lens array with a 192 $\mu$m pitch and a 3.17 mm focal length (F/16.5). The spectrograph’s backend is kept as simple as possible by using a first-order layout with identical lenses (Thorlabs TTL200MP) for the camera and collimator both having a focal length of 200 mm. The current design has diffraction-limited performance over $\pm$0.25 arcseconds on-sky, but the performance rapidly degrades outside this small field of view. The monochromatic PSFs of the spaxels can be seen in Figure \ref{fig:psflets}.

\section{Conclusion}
This manuscript has described the rational and design of a new IFU, VIS-X, for MagAO-X. It's main priority it focused on accreting proto-planets by observing H$\alpha$ at high spectral resolution. We expect that VIS-X will provide a gain in sensitivity of a 100 compared to other direct imaging instrument. This would enable us to search for fainter proto-planets at smaller angular separations. VIS-X has its first light scheduled for Fall 2021.

\acknowledgments 
The authors acknowledge funding from the Lucas/San Diego Astronomy Association Junior Faculty Award to build the VIS-X spectrograph. Support for this work was provided by NASA through the NASA Hubble Fellowship grant \#HST-HF2-51436.001-A awarded by the Space Telescope Science Institute, which is operated by the Association of Universities for Research in Astronomy, Incorporated, under NASA contract NAS5-26555. This research made use of HCIPy, an open-source object-oriented framework written in Python for performing end-to-end simulations of high-contrast imaging instruments \cite{por2018high}.

\bibliography{report} 
\bibliographystyle{spiebib} 

\end{document}